\documentclass[draft]{agujournal2019}
\usepackage{url} 
\usepackage[inline]{trackchanges} 
\usepackage{soul}

\draftfalse

\journalname{JGR: Space Physics}

\begin{document}

\title{Initial Analysis of Ionospheric Electron Density Variations Across Ecuador Using GPS Data}

\authors{E. D. López\affil{1,2}, B. A. Ubillús\affil{1}, G. L. Guamán\affil{2}}

\affiliation{1}{Escuela Politécnica Nacional, Facultad de Ciencias, Departamento de Física, Quito, Ecuador}
\affiliation{2}{Escuela Politécnica Nacional, Observatorio Astronómico de Quito/ Observatorio Nacional de Ecuador, Quito, Ecuador}

\correspondingauthor{E. D. López}{ericsson.lopez@epn.edu.ec}

\begin{keypoints}
\item TEC maps show the distribution of ionospheric electrons in Ecuador, based on GPS data from 13 strategically placed stations.
\item Temporal variations in TEC align with solar activity, with peaks occurring around local noon and decreases during the night.
\item High-resolution TEC maps provide a clearer understanding of ionospheric behavior in various regions of Ecuador.
\end{keypoints}

%
%

%
%


\begin{abstract}

In this study, we performed a preliminary mapping of Total Electron Content (TEC) over Ecuador using Global Positioning System (GPS) data. This process entails collecting and analyzing pseudorange observations from multiple GPS receivers nationwide. These receivers record signals from GPS satellites, and by comparing the arrival times of these signals, the number of electrons in the ionosphere can be inferred along the lines of sight between the satellites and the receivers.

To perform this process, signal processing algorithms are utilized to calculate TEC values, which are subsequently used to generate two-dimensional color maps that illustrate the spatial distribution of TEC in Ecuador. These maps, created using data from 13 GPS receivers distributed throughout the country, offer a valuable visualization of TEC variability regarding geographic location and time. Focusing on specific days in January 2022, this study aims to analyze patterns and trends in ionospheric electron content across the region.

The results revealed an oscillatory pattern in TEC evolution, with intensity peaks sometimes reaching or exceeding 80 TEC units (TECU), while local minima never reach zero values.
This preliminary TEC mapping approach over Ecuador using GPS data is crucial for understanding ionospheric dynamics in the region. It may have various applications, including improving the accuracy of GPS navigation, monitoring solar activity, and forecasting ionospheric phenomena that can impact communications and satellite navigation.

\end{abstract}

\section*{Plain Language Summary}

In this study, we created a basic map of the number of electrons in the ionosphere (called Total Electron Content, or TEC) over Ecuador using data from GPS. We collected information from GPS receivers placed in different parts of the country. These receivers pick up signals from GPS satellites, and by comparing when these signals arrive, we can estimate the number of electrons between the satellites and the receivers.

We used special computer methods to calculate TEC values and then created color maps to show how TEC varies across Ecuador. We focused on data from 13 GPS receivers and looked at specific days in January 2022. This helped us spot patterns in the number of electrons in the ionosphere.

The results showed that TEC changed regularly, with peaks sometimes reaching over 80 TEC units, and local lows never dropping to zero. This study is an important first step in understanding how the ionosphere behaves in Ecuador. It could help improve GPS accuracy, track solar activity, and predict ionospheric events that affect communication and satellite navigation.

%
%

%


%
%
%
%

\section{Introduction}
Global Positioning System (GPS) accuracy relies on the precise synchronization of all signal components using atomic clocks. These clocks exhibit long-term frequency stability of $10^{-13}$ to $10^{-14}$ over a day. GPS satellites are equipped with these high-precision clocks, also known as frequency standards, which generate a fundamental frequency of $10.23 \mbox{ MHz}$. From this frequency, two signals are derived: the carrier waves $f_1$ and $f_2$, obtained by multiplying the fundamental frequency by 154 and 120, respectively, resulting in $f_1 = 1575.42 \mbox{ MHz}$ and $f_2 = 1227.6 \mbox{ MHz}$ \cite{Ubillus, Hofmann}.

When radio waves, like those emitted by GPS satellites, traverse the ionosphere, they encounter two significant effects: deflection in their trajectory and a delay in their signal arrival. These effects arise from the ionosphere's free electrons, which induce a refraction effect following Snell's law \cite{Hargreaves}. However, the behavior of waves within the ionosphere is not entirely explained by this simple expression. To fully understand how radio waves behave when passing through this layer, it is crucial to consider that the ionosphere is a plasma consisting of various regions characterized by a wide range of unevenly distributed irregularities \cite{Hargreaves, Komjathy}.

Edward Appleton developed the magneto-ionospheric theory to describe the refractive index in the ionosphere. His research demonstrated that when a plane-polarized wave traverses a magnetized plasma, it separates into two circularly polarized waves rotating in opposite directions. Later, Douglas Hartree proposed incorporating Lorentz polarization into the theorem, resulting in the Appleton-Hartree formulation for calculating the complex refractive index \cite{Markovic}. The main relationships of the formulation are presented below in the methodology.


The equatorial region of the ionosphere is characterized by its high electron density. Here, the magnetic field extends horizontally, which inhibits the vertical current flow that would typically occur. This configuration leads to the accumulation of charges at the upper and lower boundaries, generating an electric field that enhances horizontal current flow. In this unique scenario, the conductivity along the magnetic field and in the horizontal direction is referred to as \textit{Cowling conductivity} \cite{Glassmeier}, which is comparable to direct conductivity, resulting in an unusually high current along the magnetic equator. This intense current is known as an equatorial "electrojet" \cite{Ericson}.


The Total Electron Content (TEC) is a key parameter of the ionosphere, representing the total number of electrons along the direct path of a signal between a receiver and a satellite \cite{Ericson, Toapanta}. Mathematically, TEC is defined as the integral of the electron density $N(s)$ along a path $ds$ between two points A and B: $TEC = \int^{B}_{A} N(s) ds$. TEC is typically measured in TECU units (1 TECU = $10^{16}$ $e$/$m^2$) and is estimated using GPS pseudorange and carrier phase data \cite{Otsuka, Scharroo}.

The behavior of TEC is fundamental in computing the ionospheric delay experienced by electromagnetic waves as they propagate through the ionosphere. Accurately predicting TEC is essential for correcting range measurements and ensuring precise satellite-based positioning. The TEC varies significantly with the amount of solar radiation received, exhibiting both diurnal (time of day) and seasonal fluctuations. Solar activity, which is linked to the sunspot cycle, also affects TEC, with increased radiation corresponding to more sunspots. Additionally, solar and magnetic disturbances impact the distribution of ionospheric plasma. These changes alter the ionosphere’s plasma density and distribution, affecting TEC and, consequently, the accuracy of satellite-based communications and navigation systems that rely on TEC for precise positioning. Understanding these phenomena is crucial for mitigating their effects on signal propagation.

In Ecuador, there has been limited research on the ionosphere and its electron content, see for instances \cite{Ubillus,Ericson,Toapanta}. As part of this series of studies, this work focuses on mapping the TEC across Ecuador using data from thirteen GPS stations collected in January 2022. By analyzing this data, the study provides insights into the temporal and spatial variability of TEC, revealing intensity fluctuations and possible seasonal patterns. Furthermore, the relationship between TEC and solar activity, including geomagnetic disturbances and coronal mass ejections, is explored to better understand external influences on ionospheric behavior. To support these analyses, visualizations such as time-series plots and geographic maps are developed, offering a clearer perspective on TEC dynamics and their distribution across the region.

The article is structured as follows: Section 2 outlines the methodology used for TEC measurements from GPS signals. Section 3 provides a detailed analysis of the mapping and analysis of the TEC. Finally, Section 4 presents conclusions drawn from the results obtained.

\section{Methodology for TEC Measurements from GPS Signals}

The methodology for TEC measurements from GPS signals begins with data collection. GPS signals from multiple satellites are acquired using ground-based GPS receivers distributed latitudinally across the country, and pseudorange and carrier phase data are recorded. Next, the slant TEC is calculated by analyzing the phase delays of the $f_1$ and $f_2$ carrier frequencies for each satellite signal, based on the Appleton-Hartree refraction model, under the approximation of disregarding the Earth's magnetic field. Afterward, a mapping function (custom code in Python) is applied to convert the slant TEC to vertical TEC, representing the total electron content along a vertical path. The TEC data is then processed to correct for errors and anomalies. Finally, the data is analyzed to study ionospheric behaviors, such as diurnal variations and ionospheric irregularities. This methodology ensures accurate TEC measurements from GPS signals, providing valuable insights into ionospheric conditions and their effects on radio wave propagation.

Let's start by highlighting some key aspects of the Appleton-Hartree equation. Firstly, it is crucial to note that this formula is applicable in an electrically neutral environment, where there are no space charges present, and there exists an equilibrium between electrons and cations. Additionally, it assumes a constant magnetic field and considers the effect of cations on the wave to be negligible \cite{Markovic, Ericson}.

To determine the ionospheric refractive index using the Appleton-Hartree equation, it is considered a plane electromagnetic wave traveling along the $x$ axis of an orthogonal coordinate system in the presence of a uniform external magnetic field that forms an angle $\theta$ with the direction of wave propagation. In this context, the complex refractive index $n$ of the Appleton-Hartree theory is expressed as follows \cite{Helliwell}:

\vspace{0.5em}

\begin{equation} \label{eb}
    n^2=1-\frac{X}{\left( 1-iZ\right)-\left[ \frac{Y^2_T}{2\left( 1-X-iZ\right)}\right]\pm \left[ \frac{Y^4_T}{4\left( 1-X-iZ\right)^2} + Y^2_L\right]^{\frac{1}{2}}},
\end{equation}

\vspace{0.5em}

\noindent with $
     X=\frac{\omega^2_N}{\omega^2} = \frac{f^2_N}{f^2}, 
     Y=\frac{\omega_H}{\omega}=\frac{f_H}{f}, 
     Y_L=\frac{\omega_L}{\omega} = \frac{f_L}{f},  Y_T=\frac{\omega_T}{\omega}=\frac{f_T}{f}, 
     Z=\frac{\omega_c}{\omega}=\frac{f_c}{f}$,

\vspace{0.5em}

\noindent where $\omega$ is the angular frequency of the carrier wave,
     $f$ is the frequency of the carrier wave with $f=\frac{\omega}{2\pi}$. The angular frequency of the plasma
     $\omega_N$ is calculated by the formula $\omega^2_N= \frac{Ne^2}{\varepsilon_0 m_e}$, with electron density $N$, electronic charge $e$, vacuum dielectric permittivity $\varepsilon_0$ and electronic mass $m_e$.
     $\omega_H$ is the cyclotron angular frequency of the free electrons, which is calculated with the formula $\omega_H = \frac{B_0|e|}{m_e}$, with magnetic induction $B_0$.
    $\omega_T$ is the transverse component of $\omega_H$, which is calculated by  $\omega_T=\frac{B_0|e|}{m_e}sin(\theta)$.
     $\omega_L$ is the longitudinal component of $\omega_H$, which is calculated by $\omega_L=\frac{B_0|e|}{m_e}cos(\theta)$ and
    $\omega_c$ is the angular frequency of collisions between electrons and heavy particles.\\

For the case in which the collisions are negligible ($Z\approx 0$), the equation \ref{eb} is reduced to $  n^2\approx 1-\frac{2X(1-X)}{2(1-X)-Y^2_T\pm \left[ Y^4_T + 4(1-X)^2 Y^2_L \right]^{\frac{1}{2}}}$ $(*)$.

\vspace{0.5em}

According to the magneto-optical theory, when a plane-polarized electromagnetic wave traverses a medium with a magnetic field, it splits into two distinct waves \cite{Hargreaves, Markovic}. The first wave, known as an \textit{ordinary wave}, behaves similarly to a wave propagating in the absence of a magnetic field. This is represented in the equation (*) with a ``+" sign. In contrast, the second wave, called an \textit{extraordinary wave}, is characterized by a ``-" sign in this equation.

Expanding equation (*) by the Taylor series, and neglecting the influence of the magnetic field ($\theta \approx 0$), we can use only the first two terms of the expansion. In this case, the refractive index of the ionospheric plasma used to calculate the TEC is simply given by:

\begin{equation}\label{e3}
    n=1-\frac{1}{2}X = 1-\frac{1}{2}\frac{f^2_N}{f^2},
\end{equation}

\vspace{0.5em}

\noindent where $f^2_N$ as a function of $N$ is $f^2_N=80.6N$ $Hz^2$. For the ionosphere, the phase refractive index suitable for carrier phase observations can be expressed as $n_p = 1-40.3 \frac{N}{f^2}$, while the group refractive index, which is appropriate for pseudorange observations, can be determined by $n_g=1+40.3 \frac{N}{f^2}$.

The delay in the propagation of electromagnetic waves, caused by the refractive index (\ref{e3}), is the primary source of error in the GPS positioning system. This delay in the arrival of the waves leads to inaccuracies in determining the exact location of the user \cite{Webster}.

\subsection{Ionospheric delay}

The delay of a signal passing through the ionosphere can be defined as the integral of the refractive index $n$ along the path $ds$ extending from the satellite ($Sat$) to the receiver ($Rec$),
$S=\int^{Rec}_{Sat} n ds$.
Replacing the refractive index $n_p$ in this equation gives:

\vspace{0.5em}

\begin{equation}\label{e7}
     S=\rho-40.3 \frac{1}{f^2}\int_{Sat}^{Rec} N ds = \rho - 40.3 \frac{TEC}{f^2},
\end{equation}

\vspace{0.5em}

\noindent where $\rho$ is the distance between the satellite and the receiver (without delays). In the last equality, the definition of the TEC  was used to replace the path integral. On the other hand, the modulated signal can be expressed equivalently as $S=\rho + 40.3 \frac{TEC}{f^2}$.

From the two previous expressions for $S$, it can be inferred that when passing through the ionosphere, the phase of the carrier wave experiences an advance (\ref{e7}) because the distance $S$ is shorter than the real distance $\rho$. Conversely, the modulated signal experiences a delay because the distance $S$ is greater than the actual distance $\rho$.
The terms subtracted and added in the respective expressions correspond to the difference between the real distance $\rho$ and the distance $S$, representing the error caused by the signal propagation through the ionosphere \cite{Webster, Garner, Ismail}. This error is known as \textit{ionospheric delay} ($d_{ion} = 40.3 ~\frac{TEC}{f^2}$).

\vspace{0.5em}

By knowing the frequencies of the GPS signal, it can be established that the delay of the ionospheric signal is exclusively dependent on the TEC. Therefore, understanding the TEC and its characteristics allows for the modeling of the ionosphere for various scientific purposes. For example, it enables the prediction of solar storms based on the development and changes in the number of electrons, as well as the determination of propagation errors in radio waves \cite{Nishioka}.

\subsection{Pseudorange-based TEC calculation}

For a specific signal of frequency $f_i$ with $i=1,2$, the pseudorange $P_i$ can be expressed using the reception time $t_r$ (measured by the receiver clock) and the transmission time $t_s$ (measured by the satellite clock) as $P_i = c \left(t_r - t_s\right)$, where $c$ is the speed of light. This equation can be rewritten for the frequencies $f_1$ and $f_2$ as:

\vspace{0.5em}

\begin{equation}\label{e10}
     P_1=\rho + c\left[ dt(t_r)-dT(t_s) \right] + d_{ion_{1}} + d_{trop} + \epsilon_{p_1},
\end{equation}

\vspace{0.5em}

\begin{equation}\label{e11}
     P_2=\rho + c\left[ dt(t_r)-dT(t_s) \right] + d_{ion_{2}} + d_{trop} + \epsilon_{p_2},
\end{equation}

\vspace{0.5em}

\noindent where $\rho$ is the range (without delays) between the satellite and the receiver,
   $c$ is the speed of light,
     $dt(t_r)$ is the receiver clock offset,
    $dT(t_s)$ is the satellite clock offset,
     $d_{ion}$ is the ionospheric delay,
      $d_{trop}$ is the tropospheric delay, and
      $\epsilon_p$ represents receiver and satellite instrumental delays, measurement noise including satellite orbital errors, and thermal noise. Subtracting the equation (\ref{e11}) from (\ref{e10}) we obtain

\vspace{0.5em}

\begin{equation}
     P_1-P_2 = d_{ion_{1}}-d_{ion_{2}} + \left( \epsilon _{p_1} - \epsilon _{p_2} \right),
\end{equation}

\vspace{0.5em}

\noindent from which the term $\left( \epsilon _{p_1} - \epsilon _{p_2} \right)$ can be neglected since its contribution to the TEC is insignificant. So, $P_1-P_2= d_{ion_{1}}-d_{ion_{2}}$. Substituting the expressions for $d_{ion_{1}}$ and $d_{ion_{2}}$ we arrive at:

\vspace{0.5em}

\[ 
   P_1-P_2= 40.3\frac{TEC_p}{f^2_1}-40.3\frac{TEC_p}{f^2_2}.
\]

\vspace{0.5em}

\noindent Consequently, the TEC of the pseudorange is calculated by:

\vspace{0.5em}

\begin{equation}\label{e12}
     TEC_p= \frac{1}{40.3}\left( \frac{f_1 f_2}{f_1-f_2} \right) \left( P_2-P_1 \right).
\end{equation}

\vspace{0.5em}

The TEC between the satellite and the user, also known as Slant Total Electron Content (STEC), varies with the satellite's elevation angle and represents the total electron density along the signal path from the satellite to the receiver \cite{Markovic, Jakowsi}.

The expression \ref{e12} is the one we use in this work to calculate the TEC, using the pseudorange data from the different receiving stations. The term involving the frequencies $f_1$ and $f_2$ is minimal in determining the TEC, whereas the difference in pseudorange is much more significant.

\subsection{GNSS Data Processing and Calibration for Total Electron Content (TEC) Analysis}

\subsubsection{Data Processing}

The \textit{Geodetic Continuous Monitoring Network of Ecuador} (REGME) generates 24-hour daily data files in its primary raw format, conforming to the \textit{RINEX 2.11} standard, with a time interval of one second. Each GPS station records pseudorange and carrier phase data at 30-second intervals. These data files are made available for download from the official \textit{Military Geographic Institute of Ecuador} (IGM) website:
\url{http://www.geograficomilitar.gob.ec/}. To collect raw data from GPS receivers, we selected 13 stations strategically distributed across the ecuadorian territory. The geographic locations of these receivers are depicted in Figure \ref{maps}. The data acquisition was focused exclusively on January 2022, a period characterized by quiet geomagnetic conditions, free from storms or disturbances. This calm month provided a stable environment, ensuring that TEC values remained consistent and regular, facilitating reliable analysis.

The raw data underwent a thorough cleaning process to remove outliers and uncertain values. Following this, we proceeded with the calculation of the STEC, a crucial step for accurately determining the ionospheric electron density and ensuring the precision of subsequent analyses.

The STEC depends on the positions of both the satellite and the GPS receiver. To derive a value that is solely dependent on the geographical location of the GPS receiver, it is necessary to project the STEC onto a vertical equivalent, known as the Vertical Total Electron Content (VTEC). This parameter represents the total number of electrons in a column perpendicular to the Earth's surface and is critical for ionospheric studies. The relationship between STEC and VTEC is given by:  

\[
STEC = MF(z, z') \times VTEC,
\]  
\noindent
where \(MF(z, z')\) is a mapping function that facilitates the transformation and \(z'\) is the zenith angle at the IPP, and it is derived from:  

\[
\sin(z') = \frac{R_E}{R_E + h_m} \sin(z),
\]  
\noindent
with \(z\) being the zenith angle at the receiver, \(R_E\) representing the Earth's average radius (6371 km), and \(h_m\) denoting the assumed height of the thin ionospheric layer (typically 350 km).

A commonly used approach employs a simple mapping function $MF= 1/\cos(z')$ to convert slant TEC values into vertical ones, associating them with specific latitudes and longitudes of the ionospheric pierce point (IPP) and the receiver. The conversion assumes that the ionosphere can be approximated as a thin spherical shell compressed at a mean altitude of approximately 350 to 450 km. This thin-layer approximation simplifies the ionosphere's complex structure, making it feasible to calculate VTEC values for a wide range of ionospheric studies, including modeling electron density variations and analyzing spatial and temporal TEC dynamics.

To ensure high-quality data for calculating VTEC, a satellite elevation angle cut-off of 30 degrees is applied. This threshold eliminates low-elevation signals, which are more susceptible to multipath effects and distortion from mapping functions (MF). At higher elevations, the effects of the troposphere on the signal are reduced, further improving measurement accuracy. Additionally, various mapping functions, such as the Klobuchar Mapping Function (KMF) and the Spherical Layer Model (SLM), produce nearly identical results at this angle, ensuring consistency and minimizing variations in VTEC calculations \cite{Schaer, Mannucci}.

\citeA{Schaer} recommended a minimum elevation angle of 10 degrees for reliable GPS signal tracking. This angle ensures sufficient signal strength while reducing atmospheric interference. As the elevation angle increases, the tropospheric influence diminishes, further enhancing the reliability of measurements. Schaer emphasized the importance of selecting appropriate elevation angles to improve signal quality and reduce errors in GPS-based applications \cite{Schaer}.

A Fortran program was used to convert the RINEX 2.11 files to compact RINEX format \cite{Rideout}. This conversion allowed for the calculation of the TEC, with the results saved in a .txt file containing two columns: the first column displaying the date and the second column showing the corresponding TEC values. Subsequently, a Python program was developed and implemented to convert the STEC into the VTEC, ensuring accurate processing of ionospheric data.

\subsubsection{Data Calibration}

Satellite and receiver biases, commonly referred to as differential code biases (DCBs), originate from the hardware of both the satellite and the receiver. These biases can be obtained from international data sources, such as the International GNSS Service (IGS), which analyze the data using specialized algorithms and software. This center produce and publish DCBs estimates for both satellites and receivers as part of their regular global ionospheric product updates. We utilized the DCBs published by the IGS to eliminate systematic biases and errors from our TEC data, guaranteeing its accuracy for analysis.

On the other hand, to address errors caused by multipath and noise, correction techniques and statistical methods are applied. Approaches such as moving averages and low-pass filters are utilized to minimize these effects, thereby enhancing data quality.

Furthermore, the data underwent a rigorous validation process. This involved cross-comparison with other reliable data sources, such as those provided by the IGS or by the regional Global Navigation Satellite System (GNSS) networks. The comparison focused on key parameters, including TEC values and temporal variations, to identify and correct discrepancies. By aligning our data with trusted external datasets, we ensured consistency and minimized systematic errors, providing independent verification of our methodology, and improving confidence in the accuracy of our results. Moreover, the TEC variations observed at one station were cross-compared with those recorded at the remaining 12 stations. The analysis revealed that TEC variations are consistent across all stations and align with expected patterns, further confirming the reliability of our data.

In an upcoming project, we plan to further compare our findings with data provided by NOAA and GPS agencies, using different GPS stations in Ecuador and over various time periods. This comparison is crucial for validating the TEC mapping data obtained in our project. Meanwhile, we recognize that discrepancies between the values reported by NOAA and IGS can be attributed to the different mapping techniques employed by each agency. NOAA uses a mapping method based on Empirical Orthogonal Functions (EOF) implemented within a Kalman filter, while IGS generates TEC maps using Global Ionosphere Maps (GIM) through their IONosphere Map EXchange (IONEX) files. In contrast, our approach focuses on analyzing the spatial and temporal variability of electron density data collected from ground-based GPS receivers distributed throughout Ecuador.

\section{Mapping and Analysis of Total Electron Content (TEC)}

To perform cartographic analysis and create TEC maps, data are processed and visualized using ArcGIS, a sophisticated Geographic Information System (GIS) software. ArcGIS provides comprehensive tools for spatial analysis, mapping, and data integration, allowing for the visualization of TEC values throughout Ecuador. This software facilitates the import of digital layers and enables the application of various geospatial analysis methods to assess the distribution of TEC throughout the region. 

To begin with, a digital map layer that highlights the provinces of Ecuador is downloaded from the Web. This layer consists of 24 distinct entities, each representing a province as a discrete geographical feature on the map. Each entity includes attribute data detailing the specific characteristics of the corresponding province. It is important to note that for the purposes of the cartographic analysis, each province is treated as an individual entity.

Before exporting the TEC data to ArcMap (a component of ArcGIS), it is essential to organize and consolidate the results, as they were initially stored in separate files for each of the 13 GPS stations. To accomplish this, a C++ program was developed to process the files for each station on a given day, extract the TEC data, and compile the information into a table within an Excel spreadsheet. Each table contains key details such as station name, latitude, longitude, date, and corresponding TEC value. For this study, four tables were generated, each representing a specific moment in time on four different days in January. These tables were then exported to ArcMap for further analysis and visualization. 

\begin{figure}[ht]
    \centering
    \includegraphics[scale=1.4]{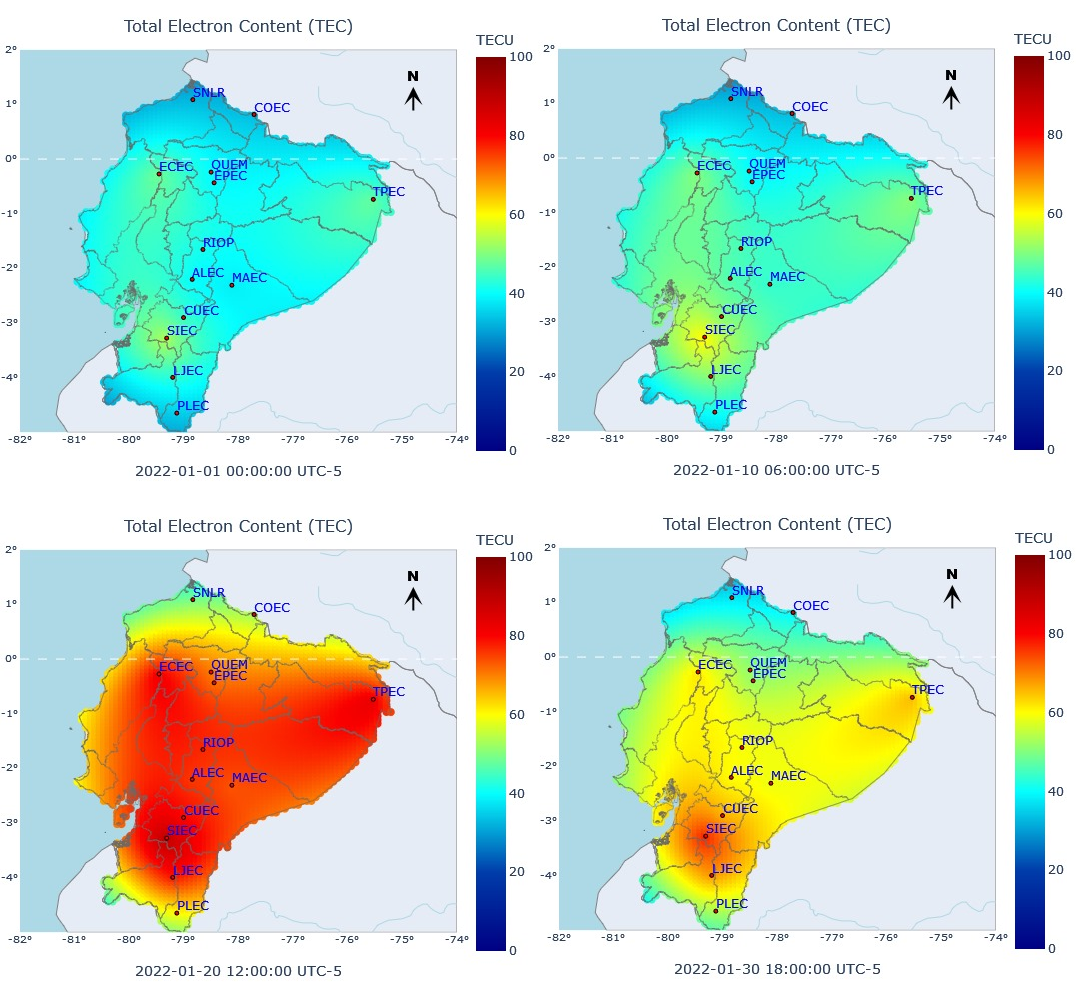}
    \caption{Maps displaying the TEC distribution over Ecuador at different hours on four distinct days in January 2022.}
    \label{maps}
\end{figure}

Once the results are exported, they are transformed into the \textit{shapefile} format, which is required for utilizing ArcMap's analytical tools. After this conversion, the Spline interpolation tool is applied by specifying the \textit{shapefile} and selecting the ``TEC" column for interpolation. This process creates a color-coded map of Ecuador that illustrates the TEC intensity, as shown in Figure \ref{maps}. Each map represents the TEC intensity at a specific moment of the day, which means the process described earlier must be repeated for three additional days within the month to generate maps for different time intervals, guaranteeing a comprehensive representation of TEC variations.

The spatial coverage of the maps spans from -5$^\circ$ to 2$^\circ$ N latitude and from -82$^\circ$ to -74$^\circ$ E longitude.

Figure \ref{maps} illustrates that the TEC reaches values between 70 and 80 TECU, which is considered high compared to other locations at higher latitudes.  Generally, TEC values tend to be higher near the equator and lower at higher latitudes. When comparing TEC values between locations, it is important to consider the specific characteristics of each location, such as latitude, altitude, and local ionospheric conditions. TEC values can vary widely based on these factors.  TEC is higher near the equator due to the stronger and more direct solar radiation that ionizes the ionosphere more efficiently. The geomagnetic field also contributes to the concentration of ionization at the equator, and the Equatorial Ionization Anomaly (EIA) creates two maxima of ionization near ±15° to ±20° latitude. In contrast, at higher latitudes, reduced solar radiation and geomagnetic effects lead to lower TEC values \cite{Hernandez2014}.

A detailed analysis of the maps shown in Figure \ref{maps} reveals distinct variations in TEC intensity throughout the day. At midnight (00:00:00), TEC values are relatively low across the entire region, reflecting minimal electron concentration. By early morning (06:00:00), a slight increase in electron density is observed in specific areas, particularly in regions with higher solar exposure. At midday (12:00:00), a substantial rise in TEC is evident across much of Ecuador, as solar radiation reaches its peak, leading to an increased ionization of the ionosphere. However, as the evening approaches (18:00:00), there is a noticeable decline in electron density, corresponding to the diminishing presence of sunlight. These temporal fluctuations in TEC are consistent with the daily cycle of solar activity, which directly influences the ionospheric electron content.

Below are time-series graphs depicting the evolution of TEC over one day, one week, and one month.

\begin{figure}[ht]
    \centering
    \includegraphics[scale=1.4]{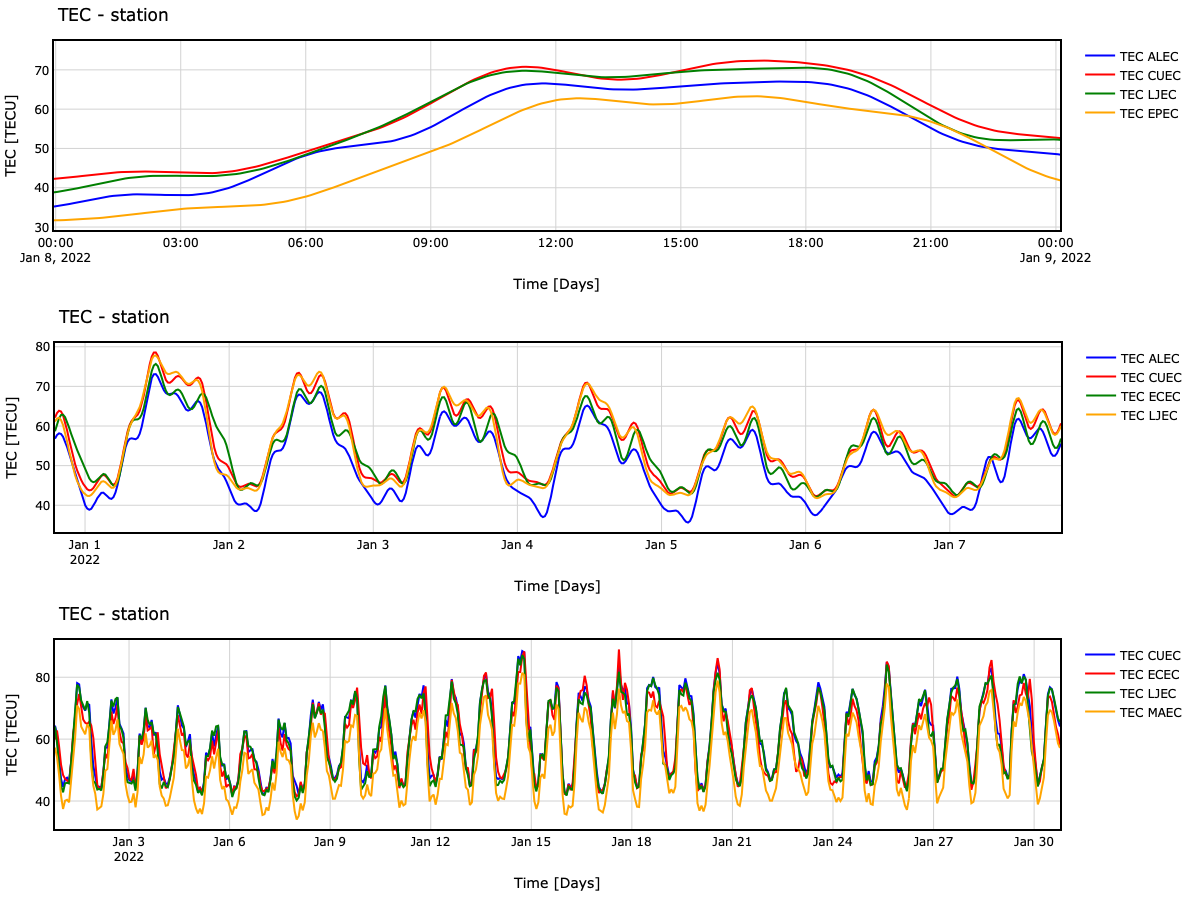}
    \caption{TEC Variation: daily (a), weekly (b), and monthly (c).}
    \label{2}
\end{figure}

The diurnal variation of TEC is depicted in Figure \ref{2}-a. It shows a minimum during the early hours of the day (00:00 UTC-5), gradually increasing in the following hours and reaching peak values between 10:00 and 16:00 UTC-5, forming a plateau, before decreasing again in the afternoon and evening. This pattern is observed at all the stations studied across Ecuador. It is primarily driven by solar radiation, which ionizes the atmosphere during the day and leads to a decrease in ionization at night. In the morning, solar radiation increases atmospheric ionization, resulting in a delayed rise in TEC. Conversely, in the evening, the reduction in solar radiation leads to decreased atmospheric ionization and a subsequent decrease in electron density.

Figure \ref{2}-a illustrates the weekly evolution of TEC, showcasing a pattern akin to the diurnal changes but occurring over a longer timeframe. This graph provides a clearer view of TEC's fluctuation over time, demonstrating an oscillatory pattern attributed to solar activity.

For the one-month interval shown in Figure \ref{2}-b, TEC oscillations resemble those of the weekly but are more pronounced.

January 2022 was chosen for this study because it was considered a quiet period in terms of geomagnetic activity, with no significant solar storms or strong magnetic disturbances . During this time, the Sun's activity was relatively calm, with no major solar events, such as solar flares or coronal mass ejections (CMEs), predicted to disrupt Earth's magnetosphere \cite{Baker2020, Liu2021}. However, contrary to these expectations, an unexpected geomagnetic disturbance developed into a G1-level storm, which resulted in the formation of a stunning aurora visible in the skies. This event was particularly notable as it was reported by observers in Preston, Lancashire, UK, who witnessed the auroral displays during the night of January 8th-9th, 2022 \cite{Smith2022}.

This unanticipated geomagnetic storm had a measurable impact on the Earth's ionosphere. As shown in Figures \ref{2}, a decrease in TEC intensity was observed across the study's measurement stations. The timing and pattern of this decrease strongly correlate with the onset of the geomagnetic disturbance, suggesting that the reduction in TEC is directly linked to the magnetic storm. The observed drop in TEC intensity likely reflects disturbances in the ionospheric layers caused by the enhanced solar wind and increased geomagnetic activity associated with the G1 storm \cite{Park2019}.

An interesting aspect of the ionosphere is that TEC values never reach zero, due to its dynamic nature. As TEC approaches near-zero levels, solar radiation begins to recharge the ionosphere at dawn, preventing complete depletion. Additionally, anomalies associated with the $F2$ region play a role in this behavior. Unlike other ionospheric regions, which may fade during the night, the $F2$ region sustains significant electron densities throughout the night, maintaining substantial levels until sunrise.

NOAA provides TEC maps in text file format, which are accessible through their website
http://www.ngdc.noaa.gov/stp/IONO/USTEC/products. The IGS (http://www.igs.org) provides files IONEX containing the global TEC maps generated by this agency. According to NOAA and IGS, in the equatorial zone and over Ecuador, the maximum values range between 70 and 90 TECU during January 2022. In Figure \ref{2} the maximum values vary between 60 and 80 TECU during January 2022, depending on the GPS station. This consistent pattern of TEC variation suggests that the evolution of TEC at each GPS station follows a similar trend, with local peaks and valleys occurring simultaneously throughout the day. Specifically, Figure \ref{2} illustrates a synchronized variation in TEC across the territory. This aspect also validates the results of this study by comparing the temporal variation of TEC values across other stations throughout the country.

We observe that the variation in TEC follows the expected pattern during the day, meaning it fluctuates depending on solar radiation incidence. There is low electron concentration at midnight, which increases with the first hours of dawn, reaches a peak during the early hours of midday and afternoon, and then decreases as the night begins. The distribution appears to be relatively uniform across the country, with minor variations. A band of lower electron concentration is observed in the north and south, which warrants further study. Interestingly, near the southernmost region of the country, in the Azuay area, high TEC values are evident throughout the daytime, in comparison with the rest of the country. Similarly, in the province of Orellana in eastern Ecuador, high TEC values are also observed, which requires detailed study that we will carry out soon to better understand their nature. Finally, along the coastal areas of the country, electron concentration is lower than in other regions of Ecuador, particularly in the inter-Andean region.

\section{Conclusions}

For the first time, comprehensive studies of TEC in the ionosphere over Ecuador are being conducted, marking a significant milestone in regional ionospheric research. These studies leverage data from GPS stations strategically distributed across the ecuadorian territory, which are fully operational, well-established, and recognized for providing high-quality data. The Military Geographic Institute of Ecuador (IGM), which oversees the majority of GNSS stations, plays a crucial role in verifying and ensuring the reliability of these datasets.

The initial TEC maps, generated for January 2022, offer an efficient and accurate depiction of TEC values and their temporal variations, along with a detailed visualization of their spatial distribution. Such granular insights are invaluable for understanding the complex behavior of the ionosphere, particularly in equatorial regions. One of the key observations from this study is that TEC values never drop to zero. This phenomenon is attributed to the continuous interaction between solar radiation and the ionosphere, even during nighttime. Additionally, the persistent anomalies in the F2 region contribute to maintaining significant electron densities after sunset. During the analysis, TEC values as high as 80 TECU were recorded at several equatorial GPS stations, emphasizing the region’s characteristic high electron density. These measurements, recorded during quiet geomagnetic conditions in January 2022, provide a clearer understanding of the ionosphere's dynamic processes in equatorial zones. The results underscore the importance of consistent monitoring and studying TEC variations, to improve our understanding of the underlying physical mechanisms and their implications for communication, navigation, and space weather forecasting. \\

Conversely, TEC maps and their time series reveal a direct correlation between incident solar energy and ionospheric electron density, resulting in predictable daily maxima and minima throughout its evolution. This pattern extends to longer time scales, where oscillatory variations in TEC are observed. These fluctuations exhibit pronounced peaks and troughs over specific periods, such as weeks and months, and are strongly influenced by solar activity. Notably, elevated electron densities were recorded in regions like Azuay (SIEC, LJEC receivers) and Orellana (TPEC receiver), highlighting the localized effects of solar-driven ionospheric dynamics.

The findings provide valuable insights into electron density distributions across the equatorial region, offering a detailed perspective on their spatial and temporal behavior. This information is critical for tracking electron movement within the ionosphere and understanding the processes that govern its dynamics.

Visual representations of TEC play a pivotal role in predicting ionospheric phenomena, such as solar storms and geomagnetic disturbances. These solar-induced events can cause significant ionospheric alterations, which, in turn, impact GPS signal propagation and other satellite-based communication systems. Continuous TEC monitoring through mapping enables the early detection of these ionospheric changes, facilitating the identification of potential space weather events and their mitigation to safeguard communication and navigation systems.

In the realm of signal propagation studies, TEC maps are invaluable for understanding how fluctuations in electron density influence electromagnetic wave behavior, particularly at frequencies ranging from 3 kHz to 30 GHz. These fluctuations can result in phenomena such as radio signal attenuation as waves traverse the ionosphere. By identifying areas of heightened attenuation, TEC maps enable proactive measures to mitigate these effects, thereby improving communication quality and enhancing GPS positioning accuracy.

In this study, we did not incorporate daytime and nighttime modulation, as the spatial variation in electron concentration was not considered. This limitation will be addressed in future work, where spatial and temporal modulation effects will be included for greater accuracy. Additionally, the thermal expansion of the atmosphere, which can indirectly influence TEC calculations, was not explicitly accounted for. While this factor is typically not included in TEC derivations from GPS measurements, it can affect the accuracy of day and night modulation corrections. This consideration remains an area for further investigation in subsequent studies.

Ongoing efforts are focused on incorporating data from approximately 50 additional GNSS stations into a central database created specifically for this research. These stations, distributed across Ecuador, cover a decade’s worth of observations, enabling a robust analysis of long-term trends in TEC behavior. Advanced computational techniques are being employed to calculate TEC values with greater precision, resulting in higher-resolution maps that capture finer details of the ionospheric variations.

\bibliography{referencias.bib}

\end{document}